\input phyzzx.tex
\tolerance=1000
\voffset=-0.0cm
\hoffset=0.7cm
\sequentialequations
\def\rl{\rightline}

\def\t1{{\tilde 1}}

\def\t{\theta}

\REF{\LAST}{E. Halyo, [arXiv:1502.01979].}
\REF{\CFTO}{M. Cadoni and S. Mingemi, Phys. Rev. {\bf D59} (1999) 081501, [arXiv:hep-th/9810251]; Nucl. Phys. {\bf B557} (1999) 165, [arXiv:hep-th/9902040].}
\REF{\CFTT}{M. Cadoni and M. Cavaglia, Phys. Lett. {\bf 499} (2001), [arXiv:hep-th/0005179]; Phys. Rev. {\bf D63} (2001) 084024,
[arXiv:hep-th/0008084].} 
\REF{\CARL}{S. Carlip, Phys. Rev. Lett. {\bf 82} (1999) 2828, [arXiv:hep-th.9812013]; Class. Quant. Grav. {\bf 16} (1999) 3327,
[arXiv:gr-qc/9906126].}
\REF{\SOL}{S. Solodukhin, Phys. Lett. {\bf B454} (1999) 213, [arXiv:hep-th/9812056].}
\REF{\GP}{A. Giacomini and N. Pinamonti, JHEP {\bf 0302} (2003) 014, [arXiv:gr-qc/0301038].}
\REF{\DL}{G. A. S. Dias and J. P. S. Lemos, Phys. Rev. {\bf D74} (2006) 044024, [arXiv:hep-th/0602144].}
\REF{\DGM}{S. Das, A. Ghosh and P. Mitra, Phys. Rev. {\bf D63} (2001) 024023,[arXiv:hep-th/0005108].}
\REF{\HSS}{M. Hotta, K. Sasaki and T. Sasaki, Class. Quant. Grav. {\bf 18} (2001) 1823, [arXiv:gr-qc/0011043].}
\REF{\MIP}{M. I. Park, Nucl. Phys. {\bf B634} (2002) 339, [arXiv:hep-th/0111224].}
\REF{\CARLI}{S. Carlip, Phys. Rev. Lett. {\bf 88} (2002) 241301, [arXiv:gr-qc//0203001].}
\REF{\DGW}{O. Dreyer, A. Ghosh and J. Wisniewski, Class. Quant. Grav. {\bf 18} (2001) 1929, [arXiv:hep-th/0101117].}
\REF{\CPP}{M. Cvitan, S. Pallua and P. Prester, Phys. Rev. {\bf D70} (2004) 084043, [arXiv:hep-th/0406186].} 
\REF{\KKP}{G. Kang, J. I. Koga and M. I. Park, Phys. Rev. {\bf D70} (2004) 024005, [arXiv:hep-th/0402113].}
\REF{\CHU}{H. Chung, Phys. Rev. {\bf D83} (2011) 084017, [arXiv:1011.0623].} 
\REF{\SIL}{S. Silva, Class. Quant. Grav. {\bf 19} (2002) 3947, [arXiv:hep-th/0204179].}
\REF{\MAJ}{B. R. Majhi and T. Padmanabhan, Phys. Rev. {\bf D85} (2012) [arXiv:1111.1809]; Phys. Rev. {\bf D86} (2012) 101501,
[arXiv:1204.1422].}
\REF{\LEN}{L. Susskind, [arXiv:hep-th/9309145].}
\REF{\SBH}{E. Halyo, A. Rajaraman and L. Susskind, Phys. Lett. {\bf B392} (1997) 319, [arXiv:hep-th/9605112].}
\REF{\HRS}{E. Halyo, B. Kol, A. Rajaraman and L. Susskind, Phys. Lett. {\bf B401} (1997) 15, [arXiv:hep-th/9609075].}
\REF{\EDI}{E. Halyo, Int. Journ. Mod. Phys. {\bf A14} (1999) 3831, [arXiv:hep-th/9610068]; Mod. Phys. Lett. {\bf A13} (1998), [arXiv:hep-th/9611175].}
\REF{\DES}{E. Halyo, [arXiv:hep-th/0107169].}
\REF{\UNI}{E. Halyo, JHEP {\bf 0112} (2001) 005, [arXiv:hep-th/0108167]; [arXiv:hep-th/0308166].}
\REF{\EDIH}{E. Halyo, [arXiv:1406.5763].}
\REF{\WAL}{R. M. Wald, Phys. Rev. {\bf D48} (1993) 3427, [arXiv:gr-gc/9307038]; V. Iyer and R. M. Wald, Phys. Rev. {\bf D50} (1994) 846, [arXiv:gr-qc/9403028]; Phys. Rev. {\bf D52} (1995) 4430, [arXiv:gr-qc/9503052].}
\REF{\EDIW}{E. Halyo. [arXiv:1403.2333].}
\REF{\TRA}{D. Birmingham, K. S. Gupta and S. Sen, Phys.Lett. {\bf B505} (2001) 191, [arXiv:hep-th/0102051]; 
A. J. M. Medved, D. Martin and M. Visser, Phys.Rev. {\bf D70} (2004) 024009 [arXiv:gr-qc/0403026].} 
\REF{\KLE}{S. Bertini, S. L. Cacciatori and D. Klemm, Phys. Rev. {\bf D85} (2012) 064018, [arXiv:1106.0999].}
\REF{\CAR}{J. L. Cardy, Nucl. Phys. {\bf B463} (1986) 435.}
\REF{\FAT}{J. Maldacena and L. Susskind, Nucl. Phys. {\bf B475} (1996) 679, [arXiv:hep-th/9604042].}
\REF{\CQM}{A. Strominger, JHEP {\bf 9901} (1999) 007, [arXiv:hep-th/9809027].}
\REF{\RIN}{M. Parikh and P. Samantray, [arXiv:1211.7370].}
\REF{\MET}{M. Cadoni and S. Mingemi, Phys. Rev. {\bf D51} (1995) 4319, [arXiv:hep-th/9410041].}
\REF{\BTZ}{A. Strominger, JHEP {\bf 9802} (1998) 009, [arXiv:hep-th/9712251].}
\REF{\BEK}{E. Halyo, JHEP {\bf 1004} (2010) 097, [arXiv:0906.2164].}
\REF{\ASH}{I. Mandal and A. Sen, Class. Quant. Grav. {\bf 27} (2010) 214003, [arXiv:1008.3801]; A. Dabholkar, Lect. Notes Phys.
{\bf 851} (2012) 165, [arXiv:1208.4814].}
\REF{\SON}{E. Halyo, in preparation.}

\singlespace
\rl{SU-ITP-15/03}
\pagenumber=0
\normalspace
\medskip
\bigskip
\titlestyle{\bf{Horizon Conformal Field Theories from $AdS_2$ Black Holes}}
\smallskip
\author{ Edi Halyo{\footnote*{e--mail address: halyo@stanford.edu}}}
\smallskip
\centerline {Department of Physics} 
\centerline{Stanford University} 
\centerline {Stanford, CA 94305}
\smallskip
\vskip 2 cm
\titlestyle{\bf Abstract}
We show that the very near horizon region of nonextreme black holes, which can be described by horizon CFTs, are related to $AdS_2$ Rindler spaces. The latter are $AdS_2$ black holes with specific masses and can be described by states of either $D=1$ or $D=2$ CFTs. The central charges of these CFTs and the conformal weights of their states that correspond to the nonextreme black holes exactly match those predicted by the horizon CFTs, providing supporting evidence for this description.

\singlespace
\vskip 0.5cm
\endpage
\normalspace

\centerline{\bf 1. Introduction}
\medskip

Recently it was shown that any nonextreme black hole in any theory of gravity can be described by a state of a $D=2$ chiral 
conformal field theory (CFT) that lives on the black hole horizon[\LAST]. More precisely, the near horizon region of a nonextreme black hole is described by the Rindler space metric, the dimensionless Rindler temperature $T_R=1/2 \pi$ and the dimensionless Rindler energy $E_R$. This can be identified with a state of conformal weight $L_0=E_R$ in a chiral $D=2$ CFT with central charge $c=12E_R$. The CFT lives in the very near horizon region, i.e. around the origin of Rindler space. 
Due to the exponential map between the Euclidean and Rindler spaces, conformal weights in Rindler space
are shifted by $-c/24=-E_R/2$. The entropy of this CFT state is given by the Cardy formula and exactly matches the Wald entropy
of the black hole.

In this paper, we provide additional evidence for the horizon CFT description of nonextreme black holes. We show that
the very near horizon geometry of the black holes, i.e. the region around the origin of Rindler space, can be mapped to $AdS_2$ Rindler space which is an $AdS_2$
black hole with a specific mass. Using the AdS/CFT correspondence, $AdS_2$ black holes can be described by states of a $D=1$ chiral
CFT that lives on the time--like boundary. Even though the details of this boundary CFT are not well--understood,
its central charge, $c$, and the conformal weights of its states, $L_0$, have been computed using the asymptotic symmetry group which is generated by a Virasoro algebra[\CFTO]. We show that for the specific black hole that corresponds to
the $AdS_2$ Rindler space in question the values of $c$ and $L_0$ exactly match those predicted by the horizon CFT. $AdS_2$ black holes can also be described by $D=2$ CFTs that are world--sheet theories of open bosonic strings[\CFTT]. In this bulk CFT, $c$ and $L_0$ of the state that corresponds to $AdS_2$ Rindler space again exactly match those predicted by the horizon CFT. We interpret these results as additional evidence that states of horizon CFTs describe nonextreme black holes.

This paper is organized as follows. In the next section, we review the description of nonextreme black holes by states of 
$D=2$ chiral CFTs that live near the black hole horizon.
In section 3, we map Rindler space to a particular $AdS_2$ black hole (i.e. the $AdS_2$ Rindler space) and show that the CFT state that describes the latter is exactly the horizon CFT state that was proposed in section 2. Section 4 contains a discussion of our results and our conclusions.

\bigskip
\centerline{\bf 2. Horizon CFTs as Nonextreme Black Holes}
\medskip

In this section, we review horizon CFTs that describe nonextreme black holes.
{\footnote1{For previous work on CFTs that describe black holes see refs. [\CARL-\MAJ].}}
It is well-known that the near horizon geometry of a nonextreme black hole in any theory of gravity is Rindler space.
Consider a black hole with a generic metric of the form
$$ ds^2=-f(r)~ dt^2+ f(r)^{-1} dr^2+ r^2 d \Omega^2_{D-2} \quad, \eqno(1)$$
in D dimensions. The horizon is at $r_h$ which is determined by
$f(r_h)=0$. If in addition, $f^{\prime}(r_h) \not =0$, the black hole is nonextreme and the near horizon geometry is described by Rindler space. Near the horizon, $r=r_h +y$ with $y<<r_h$, which leads to the near horizon metric
$$ds^2=-f^{\prime}(r_h)y~ dt^2+(f^{\prime}(r_h)y)^{-1} dy^2+ r_h^2 d \Omega^2_{D-2} \quad. \eqno(2)$$
In terms of the proper radial distance, $\rho$, obtained from $d\rho=dy/\sqrt{f^{\prime}(r_h)y}$  and the dimensionless Euclidean Rindler time $\tau=i(f^{\prime}(r_h)/2)~ t$ the near horizon metric becomes
{\footnote2{In the following all metrics will be in Euclidean signature.}}
$$ds^2=\rho^2 d \tau^2 + d \rho^2 + r_h^2 d \Omega^2_{D-2} \quad, \eqno(3)$$
where the Rindler space, i.e. the metric in the $\tau$--$\rho$ directions naively looks like the flat metric in polar coordinates.  
If we interpret the dimensionless Rindler time in eq. (3) as an angle then $E_R$ which is conjugate to $\tau$ becomes the angular
momentum of the black hole state (not to be confused with the angular momentum of the black hole in space--time).

The dimensionless Rindler energy $E_R$ conjugate to $\tau$ is obtained from the Poisson bracket[\LEN]
{\footnote3{The i on the left--hand side is due to the Euclidean signature for time.}}
$$i=\{E_R,\tau\}=\left({\partial E_R \over \partial M}{\partial \tau \over \partial t}-{\partial E_R \over \partial t}
{\partial \tau \over \partial M} \right) \quad, \eqno(4)$$
where $M$ is the mass of the black hole conjugate to $t$. For large enough black holes, the rate of Hawking radiation is negligible and therefore we can assume that $E_R$ is time independent. Then, we find 
$$dE_R={2 \over f^{\prime}(r_h)}~ dM \quad.\eqno(5)$$
Using the definition of Hawking temperature obtained from the metric, $T_H=f^{\prime}(r_h)/4 \pi$, eq. (5) can be written as
the First Law of Thermodynamics with the entropy given by $S=2 \pi E_R$. 
This procedure can be used for all nonextreme black objects with Rindler--like near horizon geometries in any theory of 
gravity[\SBH-\UNI].
In fact, it can be shown that $E_R$, which is a holographic quantity that can be obtained from a surface integral over the horizon[\EDIH], is exactly Wald's Noether charge $Q$[\WAL] and therefore[\EDIW]
$$S_{Wald}=2 \pi Q=2 \pi E_R \quad. \eqno(6)$$ 
Since the metric in eq. (3) is the same for any nonextreme black hole all the information about a particular black hole now resides 
in the dimensionless Rindler energy $E_R(M)$. 

Recently it was shown that a nonextreme black hole with a near horizon geometry that is Rindler space can be described by a state of a $D=2$ chiral CFT[\LAST]. 
More specifically, the near horizon region of a nonextreme black hole is described by Rindler space with the metric in eq. (3), the dimensionless Rindler temperature $T_R=1/2 \pi$ and the dimensionless Rindler energy $E_R$. The very near horizon region, i.e.
the region around the origin of Rindler space is described by a $D=2$ CFT since in this region all dimensionful parameters are negligible and the transverse (to $\tau$ and $\rho$) directions decouple[\CARL,\SOL,\CHU,\TRA].
{\footnote4{In addition, the scalar wave equation has an $SL(2,R)$ symmetry at low frequencies in the full Schwarzschild geometry[\KLE].}}
The horizon CFT is chiral since the Rindler metric in eq. (3) has only a $U(1)$ isometry which gets enhanced to a Virasoro algebra in the near horizon region. In addition,
we saw above that $E_R$ is both the dimensionless energy and angular momentum in Rindler space. On the other hand,
in a CFT the dimensionless Hamiltonian and angular momentum are given by
$$H=L_0+{\bar L}_0=J=L_0-{\bar L}_0=E_R \quad. \eqno(7)$$ 
Therefore, we identify Rindler space with a chiral CFT with $L_0=E_R$ and ${\bar L}_0=0$. 
We also demand that the dimensionless Rindler temperature $T_R=1/2\pi$ be equal to the dimensionless CFT temperature which satisfies
$$S={\pi^2 \over 3}cT_{CFT} \quad. \eqno(8)$$
Here $T_{CFT}$ is defined as
$$T_{CFT}={1 \over \pi} \sqrt{{{6 \Delta} \over c}} \quad, \eqno(9)$$
in order to reproduce the Cardy formula for entropy[\CAR]
$$S=2\pi \sqrt{{{c \Delta} \over 6}} \quad, \eqno(10)$$
where $\Delta$ is the conformal weight of the state in Rindler space. Rindler space is obtained from the Euclidean plane by an exponential transformation and therefore eigenvalues of $\Delta$ are shifted relative to those of $L_0$[\LAST].
This shift can be computed to be $-c/24$ and therefore, taking the Euclidean vacuum to be at zero energy, 
$\Delta=L_0-c/24$. Using $T_R=T_{CFT}=1/2 \pi$ we find that for the chiral CFT state that describes the black hole [\LAST]
$$L_0={c \over {12}}=E_R \quad. \eqno(11)$$
Plugging these values into the Cardy formula gives $S_{CFT}=2 \pi E_R=S_{Wald}$.
Therefore, we can identify the very near horizon region of a black hole with a $D=2$ chiral CFT state that satisfies eq. (11). We do not know the details of this CFT but fortunately this is not necessary
for computing the entropy through the Cardy formula. We see that black hole hair is given by momentum along the dimensionless Euclidean Rindler time direction. Unfortunately, we cannot describe the degrees of freedom that carry the hair since we do not know the details of the CFT.

We note that we do not know whether the horizon CFT is unitary and modular invariant even though we used the Cardy formula which assumes these properties. However, since the Cardy formula correctly reproduces the black hole entropy, we can turn the argument around and claim that its success is a sign that the horizon CFT must be unitary and modular invariant. In addition, the Cardy formula is only valid asymptotically for $\Delta>>c$ whereas our CFT state has $\Delta=c/24$. This problem is usually solved by invoking fractionation, i.e. by assuming that there are twisted sectors of the CFT[\FAT]. 
The dominant contribution to entropy comes from the most highly twisted sector with a twist of $E_R$. Due to the twist,
the central charge of the CFT and the conformal weight of its states are rescaled to
$c=12$ and $\Delta=E_R^2/2$ respectively. The Cardy formula can now be applied since $\Delta>>c$, and
leads to the correct black hole entropy.

\bigskip
\centerline{\bf 3. $AdS_2$ Black Holes and Rindler Space}
\medskip

We now provide additional support for the claim that the very near horizon region of a nonextreme black hole is described by a $D=2$ chiral CFT state that satisfies eq. (11). We do this by
finding a coordinate transformation from the (near horizon region of) Rindler space to $AdS_2$ Rindler space. Since these two spaces
are related by a coordinate transformation, they are classically equivalent to each other. We assume that this relation
persists quantum mechanically and therefore the CFTs that correspond to both spaces are also equivalent to each other. More precisely,
we assume that the two CFTs have the same central charge and conformal weights.
$AdS_2$ Rindler space can be described by the states of a chiral CFT that lives either on the $D=1$ boundary or the 
$D=2$ bulk. The former might be conformal quantum mechanics[\CFTO,\CQM] which is relatively well--understood. The latter is the world--sheet theory of a bosonic open string[\CFTT]. In both cases, $c$ and $L_0$ for the state that corresponds to $AdS_2$ Rindler space can be computed and exactly match those predicted by the
horizon CFT, i.e. those in eq. (11).

Consider the near horizon metric of a $D$--dimensional black hole which is Rindler space times $S^{D-2}$ given by the metric in eq. (3).
If we compactify over the horizon, i.e. over $S^{D-2}$, or equivalently consider only the s--sector of the theory by integrating out the angular directions, we obtain the two dimensional Rindler space 
$$ds^2=\rho^2 d\tau^2+d\rho^2 \quad, \eqno(12)$$
with the two--dimensional Newton constant
$G_2=G_D/A_h$ where $A_h$ is the horizon area. From this relation it is clear that the black hole entropy is given by
{\footnote5{Here we are explicitly working in General Relativity since we are assuming that entropy is given by the horizon area.}}
$$S_{BH}={A_h \over {4G_D}}={1 \over {4 G_2}} \quad, \eqno(13)$$
in two dimensional Rindler space.
Since $E_R=S_{BH}/2 \pi$, we need to show that the near horizon region of Rindler space in eq. (12) is described by a state of a $D=2$ chiral CFT with $c/12=L_0=1/8 \pi G_2$.

In order to do this we map the near--horizon region of Rindler space to $AdS_2$ which is known to be described by a chiral CFT.
Consider the Euclidean $AdS_2$ space described by the metric
$$ds^2= \left({r^2 \over \ell^2}-1 \right)dt^2+\left({r^2 \over \ell^2}-1 \right)^{-1}dr^2 \quad. \eqno(14)$$
This is actually $AdS_2$ Rindler space with an acceleration horizon at $r=\ell$. The coordinate transformation
$\rho=\sqrt{r^2-\ell^2}$ takes the metric in eq. (14) to[\RIN]
$$ds^2={\rho^2 \over \ell^2}dt^2+\left(1+{\rho^2 \over \ell^2} \right)^{-1} d\rho^2 \quad. \eqno(15)$$
It is easy to see that for $\rho<<\ell$ the metric describes Rindler space whereas for $\rho>>\ell$ it becomes that of the Poincare patch of $AdS_2$. We can map the metric in eq. (14) to Rindler space by the coordinate transformation
$${r \over \ell}= \left({{1+(\rho/2)^2} \over {1-(\rho/2)^2}} \right) \quad, \eqno(16)$$
which after the rescaling $\rho \to \ell \rho$ leads to the metric
$$ds^2={1 \over {(1-(\rho/2\ell)^2)^2}} \left({\rho^2 \over \ell^2} dt^2+d\rho^2\right) \quad. \eqno(17)$$
We see that the radius of the $AdS_2$ Rindler space, $\ell$, turns into the acceleration parameter of Rindler space, $a=1/\ell$.
Thus, $AdS_2$ Rindler space is conformal to Rindler space or more precisely to a Rindler disk since
in eq. (17) $\rho$ is bounded, $0<\rho<2 \ell$. However this is not a problem for us since we assume that all the degrees of freedom in Rindler space (at least those that contribute to the entropy) are
concentrated in the very near horizon region, i.e. in $\rho \simeq 0$. The transformation in eq. (16) maps this region to the near horizon region of $AdS_2$ Rindler space i.e. $r \simeq \ell$. We note that for $\rho<<1$, the conformal factor in eq. (17) is very close to unity and the two metrics in eqs. (14) and (17) are the same. 
{\footnote6{In a related result, ref. [\KLE] shows that the full Schwarzschild geometry can be mapped to $AdS_2$ Rindler space times $S^n$ by a Kinnersley transformation.}}
Classically, i.e. in General Relativity the two spaces described by the metrics in eqs. (14) and (17) are equivalent to each other. We assume that this equivalence persists quantum mechanically.
As a result, we expect that these two spaces are described by the same state of the same chiral CFT or at least by CFT states 
with the same $c$ and $L_0$ (which is all we need to compute the entropy).
As mentioned above, $AdS_2$ is described by both a $D=1$ boundary or a $D=2$ bulk CFT that are relatively well--understood. In both CFTs, one can compute $c$ and $L_0$ for the state that corresponds to $AdS_2$ Rindler space, i.e. a nonextreme black hole.

Consider dilatonic $AdS_2$ gravity with the action
$$A={1 \over 2} \int d^2x \eta \left(R+{2 \over \ell^2}\right) \quad, \eqno(18)$$
where $\eta$ is the dilaton field and the cosmological constant is given by $\Lambda=-2/\ell^2$. This theory has dilatonic 
black holes with the metric[\MET]
$$ds^2= \left({r^2 \over \ell^2}-{{2M\ell} \over \eta_0}\right)dt^2+\left({r^2 \over \ell^2}-{{2M\ell} \over \eta_0}\right)^{-1}dr^2
\quad, \eqno(19)$$
and the linear dilaton $\eta=\eta_0 r/\ell$. Here $M$ is the black hole mass and $\eta_0$ fixes the two--dimensional Newton constant by $\eta_0=1/8 \pi G_2$. The black hole horizon is at $r_h=(2M \ell^3/\eta_0)^{1/2}$. The mass, temperature and entropy of the black hole are given by[\MET]
$$M_{BH}={r_h^2 \over {16 \pi G_2 \ell^3}} \qquad T_{BH}={r_h \over {2 \pi \ell^2}} \qquad S_{BH}={r_h \over {4G_2 \ell}} \quad. \eqno(20)$$
The theory with the action in eq. (18) can be described either by a $D=2$ or $D=1$ CFT. Dilatonic black holes
correspond to states with computable values of $c$ and $L_0$ in these CFTs.

The $D=1$ CFT arises from the asymptotic symmetries of the $AdS_2$ space or through the AdS/CFT correspondence and lives on the 
time--like boundary of $AdS_2$. 
Near the boundary, the $SL(2,R)$ isometry of $AdS_2$ gets enhanced to a Virasoro algebra giving rise to a chiral CFT[\CFTO]. 
The central charge of this CFT and the conformal weights of its states can be computed by quantizing the 
asymptotic symmetry group of $AdS_2$. 
We can then obtain the entropy of the black hole state using the Cardy formula.  

On the $D=1$ time--like boundary of $AdS_2$ one has a chiral CFT with the central charge [\CFTO]
$${c \over 12}=\eta_0={1 \over {8 \pi G_2}} \quad. \eqno(21)$$
The ground state of the theory that satisfies $L_0=0$, is not a black hole with $M=0$ but global $AdS_2$ with a negative mass of 
$M \ell=-\eta_0/2=-c/24$. This means that, similarly to what happens with $AdS_3$ and BTZ black holes[\BTZ],
there is a shift in the values of the conformal weights and thus $\Delta=L_0-c/24$. In a supersymmetric theory, global $AdS_2$
would correspond to the Neveu--Schwarz vacuum whereas the $M=0$ black hole would be the Ramond vacuum. 
Including the shift, the conformal weight of the black hole state is given by 
$$\Delta=M \ell={r_h^2 \over {16 \pi G_2 \ell^2}} \quad. \eqno(22)$$  
Plugging in $c$ and $\Delta$ into the Cardy formula we obtain the black hole entropy in eq. (20).
Using eq. (21) we also find
$$L_0={1 \over {16 \pi G_2}} \left({r_h^2 \over \ell^2}+1\right) \quad. \eqno(23)$$ 

$AdS_2$ dilatonic gravity can also be described by a $D=2$ CFT that lives in the bulk of $AdS_2$[\CFTT]. In this case, $AdS_2$ which is basically a strip is taken to be the open bosonic string world--sheet. Then, the $D=2$ dilatonic gravity becomes the open string world--sheet theory. The action in eq. (18) can be mapped to the bosonic string world--sheet action with a one--to--one correspondence
between the world--sheet modes and the $AdS_2$ bulk modes. The theory on the $AdS_2$ boundary corresponds to a free string whereas away from the boundary the string has interactions. Dirichlet boundary conditions are imposed on the string endpoints which means that
there are no boundary degrees of freedom (such as conformal quantum mechanics). The asymptotic symmetry group of $AdS_2$ which is generated by the Virasoro algebra is exactly the conformal symmetry of the open bosonic string world--sheet. Therefore, dilaton gravity on $AdS_2$ is described by an interacting open bosonic string with Dirichlet boundary conditions.

The central charge of the bosonic string world--sheet is again given by eq. (21) whereas the eigenvalue of $\Delta$ is[\CFTT]
$$\Delta=M \ell={\pi \over 12} \alpha^{\prime} c T^2 \quad, \eqno(24)$$
where $\alpha^{\prime}=2 \pi \ell^2$ and $T=T_{BH}$ is the temperature on the world--sheet. We see that the $AdS_2$ radius, $\ell$, now plays the role of the string length. Plugging eqs. (21) and (24) into the Cardy formula we obtain the entropy of the state on the open string world--sheet
$$S=4 \pi^2 \ell \eta_0 T=S_{BH} \quad, \eqno(25)$$ 
where in the last step we used eq. (21) and $T=T_{BH}$ in eq. (20).

Now, the crucial point is to realize that the $AdS_2$ Rindler space, which is equivalent to the near horizon region of the original nonextreme black hole, is 
an $AdS_2$ black hole with $r_h=\ell$ as can be easily seen by comparing the metrics in eqs. (14) and (19). 
{\footnote6{This is exactly where the bulk Bekenstein--Hawking entropy matches the Bekenstein bound on the boundary entropy[\BEK].}}
This corresponds to a black hole with mass, i.e. $M_{BH}=1/16 \pi G_2 \ell$. 
Either eq. (23) or eq. (24) then gives $\Delta=1/16 \pi G_2$ and thus $L_0=1/8 \pi G_2$.
We conclude that $AdS_2$ Rindler space can be described by a state in either the $D=1$ or $D=2$ chiral CFT with 
$${c \over 12}=L_0=\eta_0={1 \over {8 \pi G_2}} \quad, \eqno(26)$$
and entropy $S=1/4G_2$. Since the $AdS_2$ Rindler space and Rindler space are described by the same CFT state, eq. (26) also holds
for the latter. This is precisely the horizon CFT state that we wanted to identify with Rindler space that is the near horizon
region of a nonextreme black hole.


As a consistency check, we show that we can describe the $AdS_2$ Rindler horizon by the same CFT. In order to compute $E_R$ for the metric in eq. (19) we follow the procedure outlined in section 2. For the $AdS_2$ black hole, from the metric in eq. (19) we have
$$f(r)={r^2 \over \ell^2}-{{2M \ell} \over {\eta_0}} \quad. \eqno(27)$$
Near the horizon $r=r_h+y$ with $y<<r_h$ which leads to the metric (in Euclidean signature)
$$ds^2={r_h^2 \over \ell^4} \rho^2 dt^2+ d\rho^2 \quad. \eqno(28)$$
The dimensionless Rindler energy is given by
$$E_R=\int_0^{\ell}{{\eta_0 r_h} \over {2 \ell^3}} {{2} \over (r_h/\ell^2)} dr_h=\eta_0
{r_h \over \ell}={1 \over {8 \pi G_2}}{r_h \over \ell} \quad. \eqno(29)$$
For the $AdS_2$ Rindler space $r_h=\ell$ and thus we find $E_R=1/8 \pi G_2$ which leads to exactly the same 
horizon CFT state as in eq. (26).

Above, we saw that the chirality of the horizon CFT is due to the $U(1)$ isometry of Rindler space. This single $U(1)$ can only get enhanced to a single Virasoro algebra leading to a chiral CFT. The fact that $E_R$ is both the dimensionless energy and angular momentum in Rindler space provides additional evidence for the chirality.
The relation between Rindler and and $AdS_2$ Rindler spaces also confirms the chirality of the horizon CFT.
The $D=1$ CFT on the boundary of $AdS_2$ Rindler space is chiral because of the $SL(2,R)$ isometry of $AdS_2$ which gets enhanced to
a chiral Virasoro algebra[\CFTO]. Moreover, the boundary
is one dimensional and as opposed to two dimensions, in one dimension we cannot define left or right movers. 
In terms of the $D=2$ CFT which corresponds to the open bosonic string 
world--sheet theory, chirality of the CFT arises due to the Dirichlet boundary conditions imposed on the boundary of $AdS_2$[\CFTT].

It is amusing that we found states in three chiral CFTs that describe the $AdS_2$ Rindler space with the same $c$ and $L_0$. However, the first lives on the time--like $AdS_2$ boundary, the second in the two dimensional $AdS_2$ bulk and the third on the light--like horizon. Since these three CFTs are different descriptions of nonextreme black holes,
it would be interesting to find if there are any relations between them.

\bigskip
\centerline{\bf 4. Conclusions and Discussion}
\medskip

In this paper we showed that nonextreme black holes are described by states of $D=2$ chiral CFTs
with $c/12=L_0=E_R$. We used a coordinate transformation to map (the small $\rho$ region of) Rindler space to
$AdS_2$ Rindler space. The latter has a relatively well--understood CFT description since it is a $D=2$ dilatonic black hole
with a specific mass. For the black hole state, the central charge and conformal weight, computed from the $AdS_2$ Rindler space exactly matches our horizon CFT results. In fact, there are three different CFT descriptions of $AdS_2$ Rindler spaces that correspond to nonextreme black holes: the one--dimensional asymptotic CFT that lives on the light--like $AdS_2$ boundary, the two--dimensional $AdS_2$ bulk CFT that describes the world--sheet of a bosonic open string and the CFT on the horizon of the $AdS_2$ Rindler space.
It would be interesting to find out the microscopic description of these CFTs, especially their degrees of freedom which
correspond to the hair of nonextreme black holes.

Our results are valid only for General Relativity due to our use of eq. (13) in which we explicitly assumed that the entropy of the black hole is proportional to its horizon area.  This is clearly not the case beyond General Relativity. In general, $S\not=A/4G$ 
and therefore for the Rindler space in the near horizon region of a black hole $E_R\not=1/8 \pi G_2$.
On the other hand, the relation $S_{Wald}=2 \pi E_R$ holds in any theory of gravity. In fact, in ref. [\LAST] nonextreme black holes in any theory of gravity were described by horizon CFTs.
Therefore, it should be possible to generalize our approach or find alternative one that counts Wald entropy in any theory of gravity. 

The near horizon region of a nonextreme black hole is described by only one parameter, i.e. the dimensionless Rindler energy $E_R$.
On the other hand, a CFT state is determined by two parameters: the central charge $c$ and the conformal weight $L_0$. For our description to be valid it both $c$ and $L_0$ must be fixed by $E_R$ which is certainly the case. In fact, to any CFT state with $c/12=L_0$ with large values (so that curvatures are small and gravity is classical) we can associate a Rindler space with $E_R=L_0$, i.e. the very near horizon region of a nonexteme black hole. However, this raises the question of what
deviations from the relations $c/12=L_0=E_R$ mean. For example, it would be interesting to find out whether CFT states with 
$c/12 \not= L_0$ or $c/12=L_0 \not=E_R$ describe black holes or have any other geometrical meaning.

The metric in eq. (14) that describes $AdS_2$ Rindler space is also obtained by a simultaneous near horizon and extremal limit of
nonextreme charged black holes[\ASH]. Therefore, our results can directly be applied to 
extreme charged black holes and these too can be described by the type of horizon CFTs that were the subject of this paper[\SON]. This is due to the particular near horizon and extremal limit that turns the extreme black hole with vanishing temperature into an $AdS_2$ Rindler space with nonzero temperature. As a result, $D=2$ chiral CFTs that we discussed in this paper seem to provide a uniform description of both extreme and nonextreme black hole entropy.

\endpage

\bigskip
\centerline{\bf Acknowledgments}

I would like to thank Lenny Susskind for very useful discussions and the Stanford Institute for Theoretical Physics for hospitality.

\vfill

\refout

\end
\bye